\documentclass[twocolumn,aps,amsmath,amssymb]{revtex4}

\usepackage{graphicx}
\usepackage{bm}

\begin{document}

\preprint{PRL/Ward}
\bibliographystyle{apsrev}

\title{On the physical origins of the negative index of refraction}

\author{David W. Ward}
\email{david@davidward.org} \homepage{http://www.davidward.org}
\author{Keith A. Nelson}
\affiliation{%
Department of Chemistry\\Massachusetts Institute of Technology,
Cambridge, Massachusetts 02139-4307}%
\author{Kevin J. Webb}
\affiliation{%
School of Electrical and Computer Engineering, Purdue University,
West Lafayette, IN 47907-2035}%

\date{August 23, 2004}

\begin{abstract}
The physical origins of negative refractive index are derived from
a dilute microscopic model, producing a result that is generalized
to the dense condensed phase limit. In particular, scattering from
a thin sheet of electric and magnetic dipoles driven above
resonance is used to form a fundamental description for negative
refraction. Of practical significance, loss and dispersion are
implicit in the microscopic model. While naturally occurring
negative index materials are unavailable, ferromagnetic and
ferroelectric materials provide device design opportunities.

\end{abstract}


\maketitle

In 1968, Veselago suggested that a material with a negative index
of refraction would enjoy certain peculiar properties. Foremost
among these was negative refraction, where a ray incident on the
interface of such a material would refract on the same side of the
normal rather than away from it \cite{veselago}. The subject lay
dormant until 1999, when Pendry proposed designs for magnetic
metamaterials \cite{pendry} that were subsequently implemented by
Shelby {\em et al.}, along with a dispersed electric metamaterial,
to demonstrate negative refraction in 2001 \cite{shelby}. Despite
the fact that no naturally occurring negative index material is
available, there has been a surge in interest, particularly with
regard to the prospect of creating a perfect lens \cite{pe2000}.

In this Letter, we explore the physical origins of the negative
index of refraction. The formalism developed here differs from
previous macroscopic descriptions of negative refraction in that
it is developed from a microscopic model consisting of an array of
electric dipoles and magnetic dipoles undergoing precession due to
an external static magnetic field. By considering the extension of
the microscopic model to a macroscopic electromagnetic
description, a physical basis for why the propagation speed of
light, $v_p$, should appear negative when the permittivity and
permeability are simultaneously negative is established. This
extends the macroscopic arguments originally formulated by
Veselago \cite{veselago}, and, for example, applied under the
context of a Drude model \cite{ZiHe01}, to include a physical,
microscopic basis. Finally, we use the microscopic model to
suggest some possibilities for the use of ferromagnetic and
ferroelectric materials to enhance the capabilities of negative
index metamaterials.

\begin{figure}
\includegraphics{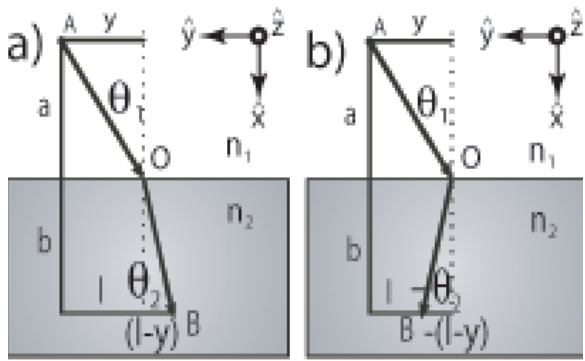}
\caption{\label{fig:leasttime} Illustration for a) positive and b) negative refraction for a light ray incident on a material with
$|n_2|>|n_1|$. In a) $0<n_1<n_2$ and in b) $0<n_1<|n_2|$ and $n_2<0$.}
\end{figure}

To illustrate how negative refraction might arise, we apply
Fermat's principle of least time \cite{born_wolf} to the case of
negative refractive index materials and interpret it in terms of
least action. Consider the path of least action for a stream of
photons in vacuum ($n_1=1$) incident on a homogeneous, isotropic
material with an index of refraction $n_2$. The most probable path
is determined by the path of stationary phase, which corresponds
to an extremum in the spatial derivative of the total travel time
through all possible paths. This is well established for the case
of positive refraction. From the diagram in
Figure~\ref{fig:leasttime}(a), the optical path length in vacuum
from source $\textbf{A}$ to the interface point $\textbf{O}$ is
$c_1t_1=\textbf{AO}=\sqrt{a^2+y^2}$, and in the semi-infinite
material from  $\textbf{O}$ to $\textbf{B}$ it is
$c_2t_2=\textbf{OB}=\sqrt{b^2+(l-y)^2}$. To find the extremum in
the time of travel from $\textbf{A}$ to $\textbf{B}$, we form
$d(t_1 + t_2)/dy=0$, with $a$ and $b$ fixed at arbitrary values,
which upon substituting for the optical path lengths gives
$n_1y/c_0\sqrt{a^2+y^2}=n_2(l-y)/c_0\sqrt{b^2+(l-y)^2}$, with $l-y
\ge 0$. Recognizing the trigonometric relations
$\sin(\theta_1)=y/\sqrt{a^2+y^2}$ and
$\sin(\theta_2)=(l-y)/\sqrt{b^2+(l-y)^2}$, this can be rewritten
in the familiar form of Snell's law:
$n_1\sin(\theta_1)=n_2\sin(\theta_2)$. If we postulate that the
photons refract to the other side of the normal, as depicted in
Figure~\ref{fig:leasttime}(b), then $l-y \le 0$, which implies
that the angle of refraction becomes $-\theta_2$. The extremum for
this optical path gives
$n_1y/c_0\sqrt{a^2+y^2}=-n_2|l-y|/c_0\sqrt{b^2+(l-y)^2}$, which
can only be satisfied if $n_2<0$. Since $n_1\sin(\theta_1)>0$ and
$\sin(-\theta_2)=-\sin(\theta_2)$, Snell's law is found to be
valid for both positive and negative refraction, provided we allow
for the possibility of a negative index of refraction. The
curvature becomes $d^2(t_1 + t_2)/dy^2 = n_1/[c_0(a^2 +
y^2)^{3/2}] + n_2/[c_0(b^2 +
  (l-y)^2)^{3/2}] $. For $n_2 > 0$, the curvature is positive,
indicating that Fermat's result indeed gives the minimum time (and
distance). Interestingly, for $n_2 < 0$, the curvature is
negative, indicating a maxima. In terms of least time this result
does not make much sense, since the path for negative refraction
corresponds a maximum, but in terms of least action, the path is
acceptable upon recognizing that negative index materials are
causal in an energy sense but not in a temporal sense, an issue we
address further in relation to a microscopic model.

The disparity between the frequency dependent speed of light in a
material and the constant speed of photons in a vacuum is resolved
by recognizing that photons impinging on a medium can drive
resonances in that medium, which may radiate and contribute to the
total scattered field. When there is a difference in phase between
the source and radiated field, the wavefronts at the detector will
appear to be advanced or retarded with respect to the source, and
it is on this basis that the concept of negative velocity is
explained.

To relate our microscopic model to the index of refraction, which
is a macroscopic phenomenon, we employ an approach developed by
Feynman, thereby extending the physical origins of the index of
refraction to include negative index materials \cite{feynman}. Our
system consists of a monochromatic plane wave source of radial
frequency $\omega$ traveling at the speed of light in vacuum,
$c_0$, in the $\hat{x}$-direction, incident on a slab of material
of thickness $d$, and a detector on the other side of the slab,
sufficiently far away. If the slab were removed, then the observed
travel time to the detector, a distance $x$ from the source, would
be commensurate with the speed of light in vacuum. When the
material is present and an observed delay or advance in arrival
time is assumed, such that the observed speed is $c_0/n$, consider
that the electric field at the detector is
\begin{equation}
E_{z}^d=E_0e^{i\frac{\omega}{c_0}d(n-1)}e^{-i\omega(t-\frac{x}{c_0})},
\label{Edetector}
\end{equation}
which is the source field of initial strength $E_0$ (neglecting
amplitude scatter in the dilute limit) multiplied by a phase term.
Feynman showed that for dilute media in which the phase term is
small, (\ref{Edetector}) can be written in a more suggestive
manner as \cite{feynman}
\begin{equation}
E_z^{d} \simeq
E_0e^{i\omega(t-\frac{x}{c_0})}-i\frac{\omega}{c_0}E_0
d(n-1)e^{i\omega(t-\frac{x}{c_0})}. \label{Edetectorthin}
\end{equation}
Written this way, it is clear in (\ref{Edetectorthin}) that the
plane wave at the detector is the sum of the source term (the
incident field without the material) and a material response term
(the scattered field). This representation is thus consistent with
the Born approximation for scattered photons.

To provide a mechanism for the apparent augmentation of the
arrival time of the photons, we employ a microscopic model that
replaces the slab with a thin sheet of electric and magnetic
dipoles, find the fields due to each source, and then apply
superposition. We assume a Lorentz electric dipole resonance and a
Bloch magnetic resonance.

The field radiated in the $x-$direction by a thin sheet of
microscopic electric dipoles is
\begin{equation}
E_z^{se}=-\frac{d}{2c_0}[i\omega \chi_\varepsilon
E_0e^{-i\omega(t-\frac{x}{c_0})}],\label{ElectricDipole}
\end{equation}
with the relative electric susceptibility given by {\cite{born}
\begin{equation}
\chi_\varepsilon=\frac{\chi_{\varepsilon_0}\omega_0^2}{(\omega_0^2-\omega^2)^2+\Gamma^2\omega^2}((\omega_0^2-\omega^2)+i\Gamma\omega),
\label{Esusceptibility}
\end{equation}
where $\chi_{\varepsilon_0}=Nq^2/m\epsilon_o\omega_0^2$ is the
static susceptibility, $N$ is the oscillator density, $q$ the
dipole charge, $\epsilon_o$ the permittivity of free space, $m$
the reduced mass of the charges in the normal mode that results in
radiation, $\omega_0$ the electric resonance frequency, and
$\Gamma$ is a phenomenological damping constant.

Also within our slab is a sheet of magnetic dipoles with a static
external field $H_0$ applied in the $\hat{x}$-direction.
Equivalently, we could assume a slab of ordered spins, where the
static magnetic field is provided by dipolar coupling, as in the
case of a ferromagnetic material. The equation for the field
radiated by a magnetic dipole sheet, $E_z^{sm}$, can be shown to
be of the same form as (\ref{ElectricDipole}) with the replacement
of $\chi_\varepsilon$ with $\chi_\mu$, the relative magnetic
susceptibility for a Bloch resonance, which is given by
\cite{slichter}
\begin{equation}
\chi_\mu=\frac{\frac{\chi_{\mu_0}}{2}\Omega_0T_2}{1+(\omega-\Omega_0)^2T_2^2}((\Omega_0-\omega)T_2+i),
\label{Msusceptibility}
\end{equation}
where $\chi_{\mu_0}$ is the ratio of the magnetization and the
static magnetic field $H_0$ that induced it, $\Omega_0$ is the
magnetic resonance frequency, and $T_2$ is the spin dephasing
time.

The field due to both the radiating electric and magnetic dipoles,
using (\ref{ElectricDipole}) and its magnetic analog, is then
\begin{equation}
E_z^s=-\frac{i\omega d}{2c_0}(\chi_\varepsilon+\chi_\mu)
E_0e^{i\omega(t-\frac{x}{c_0})} \label{dipole}.
\end{equation}
Comparing the microscopic equation of (\ref{dipole}) with our
previous macroscopic expression in (\ref{Edetectorthin}), the
index of refraction can be expressed in terms of the microscopic
susceptibilities as
\begin{equation}
n=1+\frac{1}{2}(\chi_\varepsilon+\chi_\mu). \label{nthin}
\end{equation}
We will employ some foresight and recognize that (\ref{nthin}) is
the same as the index of refraction derived from the macroscopic
Maxwell's equations for dense media of arbitrary thickness in the
limit of small susceptibilities. Accordingly, we assume the usual
form for the index of refraction, $n=\sqrt{\varepsilon_r\mu_r}$,
where the relative permittivity is $\varepsilon_r =
1+\chi_\varepsilon$ and the relative permeability is $\mu_r =
1+\chi_\mu$, so as to not limit our discussion to dilute materials
with weak susceptibilities. What we gain by doing so is that a
layer of dipoles can augment the source field for subsequent
layers of dipoles.

To describe the effect of the microscopic susceptibility on wave
propagation, we introduce the permittivity,
$\varepsilon_r=|\varepsilon_r|\exp(i\phi_\varepsilon)$, and
permeability, $\mu_r=|\mu_r|\exp(i\phi_\mu)$, in polar form into
the index of refraction ($n=
|n|e^{i\phi_n}=(|\varepsilon_r||\mu_r|)^{1/2}e^{i(\phi_\varepsilon
+\phi_\mu)/2}$), and wave admittance ($Y=
|Y|e^{i\phi_Y}=Y_0(|\varepsilon_r|/|\mu_r|)^{1/2}e^{i(\phi_\varepsilon-\phi_\mu)/2}$).
$Y_0=(\epsilon_0/\mu_0)^{1/2}$ is the free space admittance, the
inverse of the wave impedance. The domain for the index of
refraction and the wave impedance come from the range of the
permittivity and the permeability. Accordingly, with the loss
implied by the damping in (\ref{Esusceptibility}) and
(\ref{Msusceptibility}), the permittivity and permeability are
restricted to the upper-half of the complex plane, in order for a
propagating wave to lose energy (under the assumed $e^{-i \omega
t}$ convention). Thus, $\phi_{\varepsilon,\mu}$ is in the interval
$[0,\pi]$, and the index of refraction shares the same interval,
since $0\leq(\phi_\varepsilon+\phi_\mu)/2\leq\pi$. However, the
domain of the wave impedance phase is $[-\pi/2,\pi/2]$, since
$-\pi/2\leq(\phi_\mu-\phi_\varepsilon)/2\leq\pi/2$. Note that loss
becomes more significant in the neighborhood of the resonances.
However, even perturbational loss poses limitations on the
potential amplification of evanescent fields in a negative index
slab \cite{WeYaWaNe04}.

The effect of the index of refraction on the source field as it
propagates through the slab is seen by introducing the polar form
of the index of refraction into (\ref{Edetector}) and replacing
the thickness $d$ with the independent coordinate system in
Figure~\ref{fig:leasttime}. From (\ref{Edetector}), ignoring the
time dependence, and assuming $d=x$, we get $E_z^{d}(x)=E_0
e^{(-\omega x |n|/c_0)\sin\phi_n} e^{i(\omega x
|n|)/c_0)\cos\phi_n}$. By use of either the right hand rule or
Faraday's law, we find that the magnetic field associated with
propagation in the $x$-direction is $H_y^d(x)=-|Y| e^{i\phi_Y}
E_z^d (x)$. Thus, the wave admittance introduces a relative phase
shift in the electric and magnetic field, which has consequences
on power flow.

In the case of a material with only electric dipole coupling, the
value of $\phi_\varepsilon$ is restricted to the interval
$[0,\pi]$ and $\phi_\mu=0$. This limits the range of $\phi_n$ and
$\phi_Y$ to the interval $[0,\pi/2]$. However, if we allow for the
simultaneous existence of magnetic and electric dipole coupling,
we find that the interval for the phase of the index of refraction
becomes $[0,\pi]$ and the interval for the wave admittance becomes
$[-\pi/2,\pi/2]$. The most interesting result of this additional
phase contribution comes from consideration of a material that
without the magnetic dipole resonance would exhibit a bandgap for
the frequency under consideration, assuming negligible loss,
resulting in an evanescent wave. If we introduce a small magnetic
phase contribution, then we find a return to a propagating wave
solution. In the lossless electric dipole only case, the field
does not propagate and does not carry real power because the
electric and magnetic fields are $\pi/2$ radians out of phase.
However, when both electric and magnetic dipole coupling is
present, the phase contribution from the magnetic dipole can
'undo' the deleterious effects of the electric dipole by partially
restoring the relative phase of the electric and magnetic fields
such that power can flow. In the remarkable case that
$\phi_\varepsilon$ and $\phi_\mu$ are both equal to $\pi$ (or 0),
it follows that $\phi_n=\pi$ (or 0) and $\phi_Y=0$, and we find
that propagation results with no attenuation, giving a time
average Poynting vector ${\bf S}=\hat{x}\frac{1}{2}|E_0|^2
Y_0\sqrt{|\varepsilon_r|/|\mu_r|}$. If the magnitudes of the
relative permittivity and permeability are chosen to be identical,
then unimpeded propagation takes place, i.e., there is no
back-scattered wave.

The negative refractive index range, $\phi_n$ in $[\pi/2, \pi]$,
corresponds to the frequency range where $ (\chi_{\epsilon}' +
\chi_{\mu}')/2  < -1$, with $\chi_q = \chi_q' + i \chi_q^{''}$.
These overlapping frequency bands for the resonances in
(\ref{Esusceptibility}) and (\ref{Msusceptibility}) define the
frequencies where propagation can occur. Within this band, a
negative phase velocity occurs, i.e., $v_p = c_0 / n' < 0$, where
$n = n' + i n''$. This means that with advancing time, the wave
crests move in the $-x$-direction. However, in this same frequency
range, the group velocity ($v_g$), which describes the power flow
of a wave packet, is positive. This can be established from $v_g =
c/(n' + \omega {\rm d}n'/{\rm d}\omega)$ with use of
(\ref{nthin}), (\ref{Esusceptibility}), and
(\ref{Msusceptibility}). As a consequence, $(\omega/2) ( {\rm d}
\chi_{\epsilon}'/{\rm d} \omega +  {\rm d}
  \chi_{\mu}'/{\rm d} \omega) > |1 + (\chi_{\epsilon}' +
\chi_{\mu}')/2|$ in the frequency range of negative refractive
index. Therefore, conservation of power holds, i.e., with incident
power in the $x$-direction, there is power flow in the
$x$-direction in the negative refractive index medium.
Furthermore, causality can be established based on power flow,
equivalent to the procedure used in developing the Kramers-Kronig
relations for permittivity \cite{jackson}. Note that conservation
of energy is thus the basic metric for causality, and others such
as least time based on phase (velocity), for example, Fermat's
least time argument, require a revised interpretation from the
standard view for positive index materials.

The formalism developed here has direct application to
ferroelectric and ferromagnetic resonances as either the electric
dipole or the magnetic dipole source in a negative index material.
A good metric for the resonance frequency of ferromagnetics is the
Bohr magneton divided by Planck's constant, which can conveniently
be written as $\simeq 14$~GHz/Tesla \cite{slichter}. This places
the possible range of resonance frequencies somewhere between $1$
and $100$ GHz. Ferroelectrics do not have such a convenient
metric, but the resonant frequencies tend to be in $10$'s of THz
\cite{born}. Antiferromagnets can sometimes have resonance
frequencies near a THz, but they have relative susceptibilities
similar to paramagnetic materials, which are only slightly greater
than unity. It may, however, still be possible to observe a
negative index with an antiferromagnetic-ferroelectric composite.
Within their respective frequency range, ferroelectrics and
ferromagnets offer very strong resonances, with susceptibilities
approaching up to $10,000$ in both. The frequency range over which
the material may have a negative index is given by the extent of
the bandgap in the limiting material, i.e., the one with the
smallest bandgap, where the bandgap is the region between the
resonance frequency ($\omega_0$) and it's conjugate frequency
($\omega'$) . In ferroelectrics, the latter is given by the
Lyddane-Sachs-Teller relation,
$\omega'^2/\omega_0^2=\varepsilon_0/\varepsilon_{\infty}$, where
$\omega_0$ is the resonance frequency usually associated with a
transverse optical phonon in a ferroelectric crystal,
$\varepsilon_0$ is the static dielectric constant and
$\varepsilon_{\infty}$ is the high frequency permittivity usually
associated with electronic resonances \cite{born}. In
ferromagnetics, the bandgap is equal to
$\Omega_0(\mu_{\infty}-1)$, where $\mu_{\infty}$ is the relative
permeability above resonance \cite{kittel}. These materials offer
an alternative to providing both electric and magnetic resonances
with a metamaterial. Furthermore, ferromagnetic materials can be
employed now with existing metamaterials that operate normally
around $10$ GHz. One advantage in doing so is to exploit the
modest tunability of the ferromagnetic resonance and the
associated bandgap to provide a tunable negative index that could
possibly be used in a device, e.g., a switch that refracts light
one way or the other, based on the static magnetic field applied.
Ferroelectrics also offer benefits for application to existing THz
frequency split ring resonators \cite{pendry2}, in that the
electric field of an electromagnetic wave can be imaged directly
as it propagates through the ferroelectric material \cite{koehl}.

To summarize, by determining the path of stationary phase for a
stream of photons incident on an arbitrary dispersive medium, we
found that negative refraction occurs when the index of refraction
is negative, indicating that the photons traverse this path with
what seems to be a negative phase velocity. While Fermat's least
time principle correctly predicts Snell's law for refraction at an
interface with a negative refractive index material, it does not
produce a least time solution, a consequence of the negative phase
velocity. The microscopic model we have used for electric and
magnetic dipole interactions provides a basis for the collective
oscillator phase shift that results in negative phase velocity in
the case of negative refractive index. We found that when the
electric and magnetic dipoles in this model have nearly the same
resonant frequency, the augmentation of the electric component by
the Lorentz dipoles is partially compensated for by the Bloch
dipoles acting on the magnetic component, where the phase shifts
in the electric and magnetic components of the field are directly
related to the phase of the microscopic electric and magnetic
dipoles. While any simultaneous resonance in the electric and
magnetic constitutive parameters can provide a model for negative
refraction, the dilute microscopic interaction model presented,
and its generalization to the dense limit, provide a foundation
for macroscopic interpretations. As a consequence, it is clear
that dispersion and loss cannot be circumvented. Also, the dipole
resonance models presented have a finite bandgap that is
representative of physical systems such as ferroelectrics which
can, with a concomitant magnetic resonance bandgap, provide
negative refractive index over a finite frequency range. While
there are ferroelectric and ferromagnetic materials that may be
candidates for negative refractive index applications, achieving
simultaneous bandgaps in an appropriate frequency range and having
other satisfactory properties, such as low loss in the composite
system, appears challenging. On the other hand, use of one or the
other physical resonance, i.e., ferroelectric or ferromagnetic
materials, in combination with a metamaterial implementation for
the other resonance, is tractable and may yield practical
functionalities.

\bibliography{biblio}

\end{document}